# Few-View CT Reconstruction with Group-Sparsity Regularization

Peng Bao[1], Jiliu Zhou[1], Yi Zhang[1,*]

1. College of Computer Science, Sichuan University, Chengdu 610065, China

**Abstract:** Classical total variation (TV) based iterative reconstruction algorithms assume that the signal is piecewise smooth, which causes reconstruction results to suffer from the over-smoothing effect. To address this problem, this work presents a novel computed tomography (CT) reconstruction method for the few-view problem called the group-sparsity regularization-based simultaneous algebraic reconstruction technique (GSR-SART). Group-based sparse representation, which utilizes the concept of a group as the basic unit of sparse representation instead of a patch, is introduced as the image domain prior regularization term to eliminate the over-smoothing effect. By grouping the nonlocal patches into different clusters with similarity measured by Euclidean distance, the sparsity and nonlocal similarity in a single image are simultaneously explored. The split Bregman iteration algorithm is applied to obtain the numerical scheme. Experimental results demonstrate that our method both qualitatively and quantitatively outperforms several existing reconstruction methods, including filtered back projection, expectation maximization, SART, and TV-based projections onto convex sets.

## 1. Introduction

In recent decades, computed tomography (CT) has been wildly used in clinical diagnosis. However, X-ray radiation may cause cancer and genetic disease [1]. It is hence necessary to reduce the amount of a dose during a CT scan. To deal with this problem, many methods have been proposed. These methods can be categorized into two groups. The first method is to reduce the operating current, which increases the quantum noise in the projection data. The second method is to decrease the number of sampling views, which generates insufficient projection data, leading to few-view or limited-angle CT [2]. How to reconstruct a high-quality CT image from contaminated or undersampled projection data has attracted a great deal of attention in recent years. In this paper, we focus on few-view CT reconstruction.

Traditional analytic algorithms, such as filtered back projection (FBP), have specific requirements for the completeness of the projection data. Streak artifacts appear when the sampling ratio is low. The iterative reconstruction algorithm is an efficient way to solve this problem. Over the past few decades, the most widely used iterative algorithms for tomography imaging are the algebraic reconstruction technique [3], simultaneous algebraic reconstruction technique (SART) [4], and expectation maximization (EM) [5]. However, when projection views are highly sparse without extra prior information, it is very hard to obtain a satisfactory solution with these classical algorithms. To improve this problem, additional information is usually merged into the objective function to achieve a robust solution. Compressive sensing (CS) theory has been proved a powerful technique [6, 7]. If an image can be represented sparsely with a certain sparse transform, it can be accurately reconstructed with a probability close to one. Inspired by CS theory, Sidky *et al.* introduced total variation (TV) minimization into incomplete projection data reconstruction and proposed an efficient iterative reconstruction algorithm based on projection onto convex sets (POCS), called TV-POCS [8]. Although TV-POCS can eliminate streak artifacts to a certain degree, the assumption of TV that the signal is piecewise smooth causes TV-POCS to suffer from over-smoothing effects [9]. As a result, many variants of TV have been proposed to tackle this problem, such as adaptive-weighted TV [10], fractional-order TV [11, 12], and nonlocal means [13, 14]. Chen *et al.* suggested that a high-quality image can be utilized to constrain the CS-based reconstruction [15], and this method has been extended to several different reconstruction topics with different prior images. Yu *et al.* constructed the pseudo-inverses of the discrete gradient and discrete difference transforms and adopted a soft-threshold filtering algorithm for few-view CT image reconstruction [16].

Recently, dictionary learning based methods have been proved effective. In contrast to traditional techniques, which process the image pixel by pixel, a dictionary-based method processes images patch by patch. In 2006, Elad tackled the image denoising problem with a dictionary learning method that utilizes the K-SVD algorithm [17]. Mairal *et al.* extended this method to colour image restoration [18]. For medical imaging problems, the dictionary learning method was first introduced into magnetic resonance imaging (MRI). Chen *et al.* combined the dictionary learning method and TV-based MRI scheme to further improve image quality [19]. Later, Xu *et al.* proposed a low-dose CT image reconstruction method based on dictionary learning. This model introduces the sparse representation constraint of a redundant dictionary as the regularization term, and the performance of a global dictionary and adaptive dictionary was discussed [20]. Inspired by work combining super-resolution with dual dictionary learning [21], Lu *et al.* respectively used a transitional dictionary for atom matching and a global dictionary for image updating to deal with the few-view problem [22]. Zhao *et al.* extended this method for spectral CT [23].

Traditional studies based on dictionary learning have two limits. First, the computational burden is very heavy. Second, the relationships among patches are ignored. If the original signals are noisy, the accuracy of sparse coding will decline. Inspired by the research on group sparsity [24–27], in this article, we proposed a novel few-view CT reconstruction method based on group-sparsity regularization (GSR) called the GSR-based simultaneous algebraic reconstruction technique (GSR-SART). Instead of processing the image patches sequentially, similar patches are clustered into groups as the basic unit of the proposed group-based sparse representation. Thus, the sparsity and nonlocal similarity in a single image are simultaneously imposed. The remainder of the paper is organized as follows: Section 2 introduces the theory details and numerical scheme for the proposed GSR-SART. Experimental results are provided in Section 3 to demonstrate the performance of our method. A discussion and the conclusion are presented in Section 4.

## 2. Methods

### *2.1. Imaging model*

Assuming a monochromatic source, the general model of CT imaging can be approximately represented as the following discrete linear system:

$$\boldsymbol{g} = \boldsymbol{A}\boldsymbol{u}, \tag{1}$$

where $\boldsymbol{A}$ is the system matrix, which is composed of $M$ row vectors, and $\boldsymbol{g}$ denotes the measured projection data. Our goal is to reconstruct an image represented by vector $\boldsymbol{u}$ from the projection data $\boldsymbol{g}$ and system matrix $\boldsymbol{A}$. In practice, Eq. (1) is known as an ill-posed problem because we consider the insufficient projection data problem caused by few views. This means that we cannot obtain a unique $\boldsymbol{u}$ by directly inverting Eq. (1). To solve the linear system expressed in Eq. (1), prior information about the target image is often imposed.

### *2.2. Sparse Representation Modelling*

Sparse representation (SR) for image processing seeks a sparse matrix that contains as few zero coefficients as possible to approximately represent the signal. The SR model can be expressed as

$$\{\alpha, D\} = argmin_{\boldsymbol{\alpha},\boldsymbol{D}} \|\boldsymbol{x} - \boldsymbol{D}\boldsymbol{\alpha}\|_2^2 + \lambda \|\boldsymbol{\alpha}\|_0, \quad (2)$$

where $\boldsymbol{x}$ denotes an observed signal vector, $\boldsymbol{D}$ is a dictionary, α presents the coefficients to represent the signal, $\lambda$ is a regularization parameter, and $\|\cdot\|_0$ denotes the L0 norm. The goal of SR is to seek a sparse vector $\boldsymbol{\alpha}$ to represent $\boldsymbol{x}$ for a trained $\boldsymbol{D}$. To better represent $\boldsymbol{x}$ with $\boldsymbol{\alpha}$, it is necessary to choose an effective dictionary $\boldsymbol{D}$. Some approximation algorithms have been proposed to alternatively optimize $\boldsymbol{D}$ and $\boldsymbol{\alpha}$, such as MOD [28], K-SVD [29], and online learning [30].

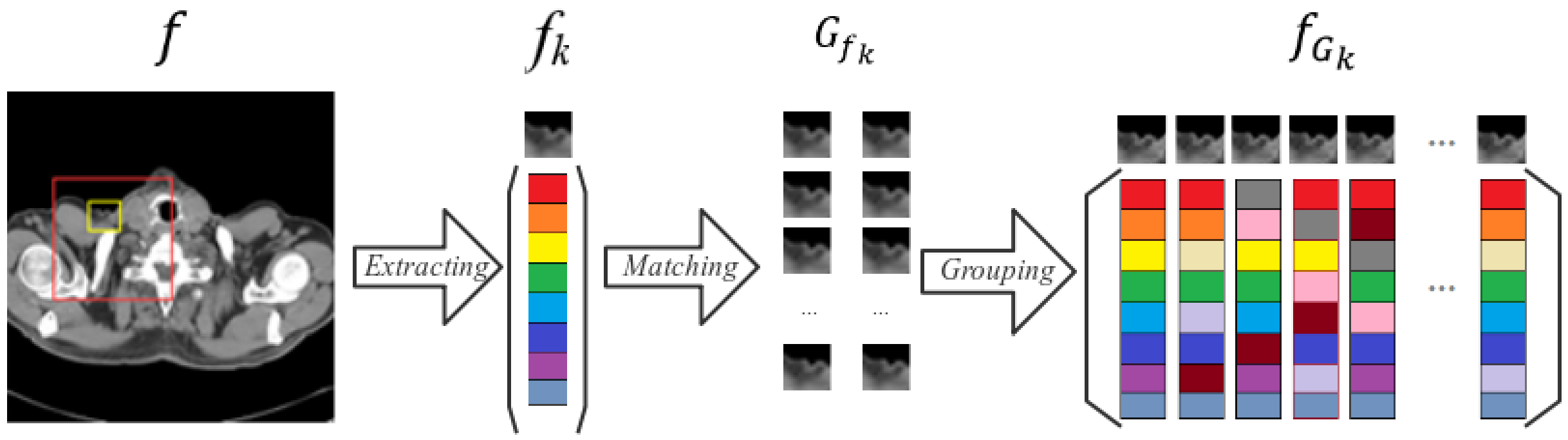

Fig. 1. Group construction. Each patch vector $\boldsymbol{f}_k$ is extracted from image $\boldsymbol{f}$. Here, $\boldsymbol{G}_{\boldsymbol{f}_k}$ denotes the set composed of $m$ most similar patches and $\boldsymbol{f}_{\boldsymbol{G}_k}$ is a matrix composed of all the patch vectors in $\boldsymbol{G}_{\boldsymbol{f}_k}$.

### *2.3. Group-Based Sparse Representation Modelling*

Usually, most SR-based methods divide the image into overlapped patches and process them one by one. This operation ignores the nonlocal relationships between different patches. In this paper, we impose the nonlocal similarity constraint into SR to create a GSR-based few-view CT reconstruction method.

First, we divide the CT image $\boldsymbol{f}$ into $n$ overlapped patches of size $\sqrt{P_s} \times \sqrt{P_s}$ using a sliding distance of four pixels, where vector $\boldsymbol{f}_k$ denotes an image patch at location $k$, $k = 1, 2, 3 \dots, n$. In Fig. 1, $\boldsymbol{f}_k$ is indicated by a small yellow square. The Euclidean distance is utilized as the similarity measurement to search for the $m$ patches that are the most similar with $\boldsymbol{f}_k$ in the $L \times L$ search window, as indicated by the big red square in Fig. 1. These similar patches form set $\boldsymbol{G}_{\boldsymbol{f}_k}$. Second, all the patches in $\boldsymbol{G}_{\boldsymbol{f}_k}$ are unfolded into vectors and arranged into a matrix of size $P_s \times m$, denoted by $\boldsymbol{f}_{\boldsymbol{G}_k}$, which includes each patch in $\boldsymbol{G}_{\boldsymbol{f}_k}$ as its columns. Matrix $\boldsymbol{f}_{\boldsymbol{G}_k}$ is treated as a group of similar patches. We can then define

$$\boldsymbol{f}_{\boldsymbol{G}_k} = \boldsymbol{E}_{\boldsymbol{G}_k}(\boldsymbol{f}), \quad (3)$$

where $\boldsymbol{E}_{\boldsymbol{G}_k}$is an operator that extracts group $\boldsymbol{f}_{\boldsymbol{G}_k}$ from $\boldsymbol{f}$. Next, we use $\boldsymbol{E}_{\boldsymbol{G}_k}^T(\boldsymbol{f}_{\boldsymbol{G}_k})$ to denote placing group $\boldsymbol{f}_{\boldsymbol{G}_k}$ back into the $k^{th}$ position of the reconstructed image. Now, we can express the whole image $\boldsymbol{f}$ by averaging all the groups as follows:

$$\boldsymbol{f} = \sum_{k=1}^{n} \boldsymbol{E}_{\boldsymbol{G}_k}^T(\boldsymbol{f}_{\boldsymbol{G}_k}) \,/\, \sum_{k=1}^{n} \boldsymbol{E}_{\boldsymbol{G}_k}^T(\boldsymbol{1}_{P_s*m}), \tag{4}$$

where operator / indicates the element-wise division of two vectors and $\boldsymbol{1}_{P_s*m}$ is an all-ones matrix of the same size as $\boldsymbol{f}_{\boldsymbol{G}_k}$.

Next, we introduce the dictionary learning method for each group $\boldsymbol{f}_{\boldsymbol{G}_k}$ . In this model, the adaptive dictionary $\boldsymbol{D}_{\boldsymbol{G}_k}$ for each group $\boldsymbol{f}_{\boldsymbol{G}_k}$ can be directly obtained from its estimate $\boldsymbol{e}_{\boldsymbol{G}_k}$ because we cannot obtain the original image $\boldsymbol{f}$ in practice. In the process of optimization, estimate $\boldsymbol{e}_{\boldsymbol{G}_k}$ is calculated. Once we obtain $\boldsymbol{e}_{\boldsymbol{G}_k}$, we apply SVD to it as follows:

$$\boldsymbol{e}_{\boldsymbol{G}_k} = \boldsymbol{U}_{\boldsymbol{G}_k}\boldsymbol{\Sigma}_{\boldsymbol{G}_k}\boldsymbol{V}_{\boldsymbol{G}_k}^T = \sum_{i=1}^{c} \boldsymbol{\beta}_{e_{\boldsymbol{G}_k*i}}(\boldsymbol{u}_{\boldsymbol{G}_k*i}\,\boldsymbol{v}_{\boldsymbol{G}_k*i}^T), \tag{5}$$

where $c$ is the number of atoms in $\boldsymbol{D}_{\boldsymbol{G}_k}$, $\boldsymbol{\beta}_{e_{\boldsymbol{G}_k}} = \{\boldsymbol{\beta}_{e_{\boldsymbol{G}_k*1}}, \boldsymbol{\beta}_{e_{\boldsymbol{G}_k*2}}, \dots, \boldsymbol{\beta}_{e_{\boldsymbol{G}_k*c}}\}$, $\boldsymbol{\Sigma}_{\boldsymbol{G}_k} = diag(\boldsymbol{\beta}_{e_{\boldsymbol{G}_k}})$ denotes a diagonal matrix for which all the elements except for the main diagonal are zero, $\boldsymbol{u}_{\boldsymbol{G}_k*i}$ denotes the columns of $\boldsymbol{U}_{\boldsymbol{G}_k}$, and $\boldsymbol{v}_{\boldsymbol{G}_k*i}^T$ denotes the columns of $\boldsymbol{V}_{\boldsymbol{G}_k}^T$. For group $\boldsymbol{f}_{\boldsymbol{G}_k}$, each atom of $\boldsymbol{D}_{\boldsymbol{G}_k}$ is defined as follows:

$$\boldsymbol{d}_{\boldsymbol{G}_k*i} = \boldsymbol{u}_{\boldsymbol{G}_k*i}\boldsymbol{v}_{\boldsymbol{G}_k*i}^T,\ i = 1,2,\dots,c \tag{6}$$

We can then define the expression of the ultimate dictionary $\boldsymbol{D}_{\boldsymbol{G}_k}$ for group $\boldsymbol{f}_{\boldsymbol{G}_k}$ as follows:

$$\boldsymbol{D}_{\boldsymbol{G}_k} = \{\boldsymbol{d}_{\boldsymbol{G}_k*1}, \boldsymbol{d}_{\boldsymbol{G}_k*2}, \dots, \boldsymbol{d}_{\boldsymbol{G}_k*c}\} \tag{7}$$

Using $\boldsymbol{D}_{\boldsymbol{G}_k}$, the GSR model seeks a sparse vector $\boldsymbol{\alpha}_{\boldsymbol{G}_k} = \{\boldsymbol{\alpha}_{\boldsymbol{G}_k*1}, \boldsymbol{\alpha}_{\boldsymbol{G}_k*2}, \dots, \boldsymbol{\alpha}_{\boldsymbol{G}_k*c}\}$ to represent $\boldsymbol{f}_{\boldsymbol{G}_k}$:

$$\boldsymbol{f}_{\boldsymbol{G}_k} = \sum_{i=1}^{c} \boldsymbol{\alpha}_{\boldsymbol{G}_k*i}\boldsymbol{d}_{\boldsymbol{G}_k*i} \tag{8}$$

We can then represent the entire image $\boldsymbol{f}$ using the set of sparse codes $\{\boldsymbol{\alpha}_{\boldsymbol{G}_k}\}$.

$$\boldsymbol{f} = \boldsymbol{D}_{\boldsymbol{G}} * \boldsymbol{\alpha}_{\boldsymbol{G}} = \sum_{k=1}^{n} \boldsymbol{E}_{\boldsymbol{G}_k}^T\left(\sum_{i=1}^{c} \boldsymbol{\alpha}_{\boldsymbol{G}_k*i}\boldsymbol{d}_{\boldsymbol{G}_k*i}\right) ./ \sum_{k=1}^{n} \boldsymbol{E}_{\boldsymbol{G}_k}^T(\boldsymbol{1}_{P_s*m}) \tag{9}$$

Here, $\boldsymbol{D}_{\boldsymbol{G}}$ represents the concatenation of all $\boldsymbol{D}_{\boldsymbol{G}_k}$ and $\boldsymbol{\alpha}_{\boldsymbol{G}}$ denotes the concatenation of all $\boldsymbol{\alpha}_{\boldsymbol{G}_k}$.

### *2.4. GSR-SART algorithm*

Similar to [31], two independent steps are included in our algorithm. In the first step, we adopt SART to solve the linear system of Eq. (1), which yields a noisy result by minimizing the distance between the measured projection data and the estimated projection data. Specifically, the SART algorithm can be described as follows:

$$\boldsymbol{u}_j^{p+1} = \boldsymbol{u}_j^p + \frac{\omega}{A_{+,j}} \sum_{i=1}^{M} \frac{\boldsymbol{A}_{i,j}}{\boldsymbol{A}_{i,+}} (\boldsymbol{g}_i - \overline{\boldsymbol{g}_i}(\boldsymbol{u}^p)), \tag{10}$$

$$\boldsymbol{A}_{i,+} = \sum_{j=1}^{N} \boldsymbol{A}_{i,j} \ \ for\ i = 1,2, \dots, M, \tag{11}$$

$$\boldsymbol{A}_{+,j} = \sum_{i=1}^{M} \boldsymbol{A}_{i,j} \ \ for\ j = 1,2, \dots, N, \tag{12}$$

$$\overline{\boldsymbol{g}}(\boldsymbol{u}) = \boldsymbol{A}\boldsymbol{u}, \tag{13}$$

where $\boldsymbol{A}$ is a system matrix of size $M \times N$ ($M$ is the total number of projection data and $N$ is the total number of image pixels), $\omega$ is the relaxation parameter, and $p$ is the iteration number. The second step is to obtain an artifact-reduced result using GSR with the estimated result $\boldsymbol{u}$ from SART as an initial value. The optimization problem of the second step can be expressed as

$$min_{\boldsymbol{D_G}, \boldsymbol{\alpha_G}, \boldsymbol{f}} \frac{1}{2} \|\boldsymbol{f} - \boldsymbol{u}\|_2^2 + \lambda \|\boldsymbol{\alpha_G}\|_0 \ \ s.t.\ \boldsymbol{f} = \boldsymbol{D_G} * \boldsymbol{\alpha_G}, \tag{14}$$

where $\frac{1}{2}\|\boldsymbol{f} - \boldsymbol{u}\|_2^2$ is an $l_2$ data-fidelity term, $\|\boldsymbol{\alpha_G}\|$ is a regularization term, and $\lambda$ is a regularization parameter. We can obtain adaptive dictionary $\boldsymbol{D_G}$ by applying SVD to the estimate $\boldsymbol{e}$ of $\boldsymbol{f}$ according Eqs. (5)–(7). How to calculate $\boldsymbol{e}$ is given below. Then, Eq. (14) becomes

$$min_{\boldsymbol{\alpha_G}, \boldsymbol{f}} \frac{1}{2} \|\boldsymbol{f} - \boldsymbol{u}\|_2^2 + \lambda \|\boldsymbol{\alpha_G}\|_0 \ \ s.t.\ \boldsymbol{f} = \boldsymbol{D_G} * \boldsymbol{\alpha_G}. \tag{15}$$

However, Eq. (15) is always hard to solve because the $l_0$-norm optimization is non-convex. In this paper, the split Bregman iteration (SBI) algorithm [32] is used to solve this problem. Consider the following constrained optimization problem:

$$min_{\boldsymbol{x}, \boldsymbol{y}} p(\boldsymbol{x}) + q(\boldsymbol{y}) \ \ s.t.\ \boldsymbol{x} = \boldsymbol{G}\boldsymbol{y}. \tag{16}$$

According to SBI, the minimization problem in Eq. (16) can be split to sub-problems, as shown in Algorithm 1.

Algorithm 1: Split Bregman Iteration (SBI)

1. Set $t = 0$, choose $\mu > 0$, $\boldsymbol{b}^0 = 0, \boldsymbol{x}^0 = 0, \boldsymbol{y}^0 = 0$
2. Repeat
3. $\boldsymbol{x}^{t+1} = argmin_{\boldsymbol{x}} p(\boldsymbol{x}) + \frac{\mu}{2} \|\boldsymbol{x} - \boldsymbol{G}\boldsymbol{y}^t - \boldsymbol{b}^t\|_2^2$
4. $\boldsymbol{y}^{t+1} = argmin_{\boldsymbol{y}} q(\boldsymbol{y}) + \frac{\mu}{2} \|\boldsymbol{x}^{t+1} - \boldsymbol{G}\boldsymbol{y} - \boldsymbol{b}^t\|_2^2$
5. $\boldsymbol{b}^{t+1} = \boldsymbol{b}^t - (\boldsymbol{x}^{t+1} - \boldsymbol{G}\boldsymbol{y}^{t+1})$
6. $t = t + 1$
7. Until the stopping criterion is satisfied

We define $p(\boldsymbol{f}) = \frac{1}{2} \|\boldsymbol{f} - \boldsymbol{u}\|_2^2, q(\boldsymbol{\alpha_G}) = \lambda \|\boldsymbol{\alpha_G}\|_0$. Then, invoking SBI, step 3 of SBI becomes:

$$\boldsymbol{f}^{t+1} = argmin_{\boldsymbol{f}} \frac{1}{2}\|\boldsymbol{f}-\boldsymbol{u}\|_2^2 + \frac{\mu}{2}\|\boldsymbol{f}-\boldsymbol{D_G} * \boldsymbol{\alpha_G^t} - \boldsymbol{b}^t\|_2^2. \tag{17}$$

Step 4 of SBI becomes:

$$\boldsymbol{\alpha_G^{t+1}} = argmin_{\boldsymbol{\alpha_G}} \lambda\|\boldsymbol{\alpha_G}\|_0 + \frac{\mu}{2}\|\boldsymbol{f}^{t+1}-\boldsymbol{D_G} * \boldsymbol{\alpha_G} - \boldsymbol{b}^t\|_2^2. \tag{18}$$

Step 5 of SBI becomes:

$$\boldsymbol{b}^{t+1} = \boldsymbol{b}^t - (\boldsymbol{f}^{t+1} - \boldsymbol{D_G} * \boldsymbol{\alpha_G^{t+1}}). \tag{19}$$

Then, the minimization of Eq. (15) is transformed into two sub-problems concerning $\boldsymbol{f}$ and $\boldsymbol{\alpha_G}$. For a given $\boldsymbol{\alpha_G}$, the $\boldsymbol{f}$ sub-problem in Eq. (17) is a strictly quadratic convex optimization problem, which can be defined as

$$min_{\boldsymbol{f}} P_1(\boldsymbol{f}) = min_{\boldsymbol{f}} \frac{1}{2}\|\boldsymbol{f}-\boldsymbol{u}\|_2^2 + \frac{\mu}{2}\|\boldsymbol{f}-\boldsymbol{D_G} * \boldsymbol{\alpha_G} - \boldsymbol{b}^t\|_2^2. \tag{20}$$

We can obtain a closed solution for Eq. (20) by setting the gradient of $P_1(\boldsymbol{f})$ to zero, which can be expressed as

$$\hat{\boldsymbol{f}} = ((1+\mu)\boldsymbol{E})^{-1}(\boldsymbol{u} + \mu(\boldsymbol{D_G} * \boldsymbol{\alpha_G} + \boldsymbol{b})), \tag{21}$$

where $\boldsymbol{E}$ is an identity matrix. For a given $\boldsymbol{f}$, the $\boldsymbol{\alpha_G}$ subproblems can be defined as

$$min_{\boldsymbol{\alpha_G}} P_2(\boldsymbol{\alpha_G}) = min_{\boldsymbol{\alpha_G}} \frac{\lambda}{\mu}\|\boldsymbol{\alpha_G}\|_0 + \frac{1}{2}\|\boldsymbol{e}-\boldsymbol{D_G} * \boldsymbol{\alpha_G}\|_2^2, \tag{22}$$

where $\boldsymbol{e} = \boldsymbol{f} - \boldsymbol{b}$. According to the theorem in [24], Eq. (22) can be transformed into

$$min_{\boldsymbol{\alpha_G}} P_2(\boldsymbol{\alpha_G}) = min_{\boldsymbol{\alpha_G}} \sum_{k=1}^{n}(\frac{1}{2}\left\|\boldsymbol{f}_{\boldsymbol{G}_k} - \boldsymbol{e}_{\boldsymbol{G}_k}\right\|_F^2 + \xi\left\|\boldsymbol{\alpha}_{\boldsymbol{G}_k}\right\|_0), \tag{23}$$

where $\xi = (\lambda \times Ps \times m \times n) / (\mu \times N)$. Then, Eq. (22) transforms into $n$ sub-problems for all groups $\boldsymbol{f}_{\boldsymbol{G}_k}$. Because $\boldsymbol{f}_{\boldsymbol{G}_k} = \boldsymbol{D}_{\boldsymbol{G}_k}\boldsymbol{\alpha}_{\boldsymbol{G}_k}$ and $\boldsymbol{e}_{\boldsymbol{G}_k} = \boldsymbol{D}_{\boldsymbol{G}_k}\boldsymbol{\beta}_{\boldsymbol{G}_k}$, the minimization for each group $\boldsymbol{f}_{\boldsymbol{G}_k}$ can be defined as follows:

$$argmin_{\boldsymbol{\alpha}_{\boldsymbol{G}_k}} \frac{1}{2}\left\|\boldsymbol{\alpha}_{\boldsymbol{G}_k} - \boldsymbol{\beta}_{\boldsymbol{G}_k}\right\|_2^2 + \xi\left\|\boldsymbol{\alpha}_{\boldsymbol{G}_k}\right\|_0. \tag{24}$$

Therefore, a closed solution for Eq. (24) can be expressed as follows:

$$\hat{\boldsymbol{\alpha}}_{\boldsymbol{G}_k} = \boldsymbol{\beta}_{\boldsymbol{e}_{\boldsymbol{G}_k}} * 1(abs\left(\boldsymbol{\beta}_{\boldsymbol{e}_{\boldsymbol{G}_k}}\right) - \sqrt{2\xi}), \tag{25}$$

where operator $*$ indicates the element-wise product between two vectors and $1(abs\left(\boldsymbol{\beta}_{\boldsymbol{e}_{\boldsymbol{G}_k}}\right) - \sqrt{2\xi})$ is defined as

$$1(abs\left(\boldsymbol{\beta}_{\boldsymbol{e}_{\boldsymbol{G}_k}}\right) - \sqrt{2\xi}) = \begin{cases} 1, & abs\left(\beta_{e_{G_k}}\right) > \sqrt{2\xi} \\ 0, & abs\left(\beta_{e_{G_k}}\right) \leq \sqrt{2\xi} \end{cases}. \tag{26}$$

Once $\boldsymbol{\alpha}_{\boldsymbol{G}_k}$ is calculated for all groups, the final solution for the $\boldsymbol{\alpha_G}$ sub-problem is determined.

### *2.5. Summary of the Proposed Algorithm*

Our algorithm is composed of two main parts: the SART reconstruction and GSR regularization. We summarize the pseudo-code of our GSR-SART algorithm in Algorithm 2.

Algorithm 2: GSR-SART

**Initialization:**

Given $\boldsymbol{u}^0, \omega, \boldsymbol{b}^0, \boldsymbol{f}^0, \boldsymbol{\alpha}_{\boldsymbol{G}}^0, \boldsymbol{D}_{\boldsymbol{G}}^0, \lambda, \mu, P_s, L, m, \xi$

**Repeat:**

**SART step:**

for $p = 1, 2, 3, \ldots, P$

if $p == 1$

$\boldsymbol{u}^p = SART\ (\boldsymbol{u}^0, \omega)$

else

$\boldsymbol{u}^p = SART\ (\boldsymbol{u}^{p-1}, \omega)$

end if

end for

if $\boldsymbol{u}^p < 0$

$\boldsymbol{u}^p = 0$

$\boldsymbol{u} = \boldsymbol{u}^p$

**GSR:**

for $t = 0, 1, 2, \ldots, T$

update $\boldsymbol{f}^{t+1}$ by Eq. (21)

$\boldsymbol{e}^{t+1} = \boldsymbol{f}^{t+1} - \boldsymbol{b}^t$

for each group $\boldsymbol{f}_{\boldsymbol{G}_k}$

update $\boldsymbol{D}_{\boldsymbol{G}_k}$ by Eq. (5) - (7)

update $\boldsymbol{\beta}_{\boldsymbol{e}_{\boldsymbol{G}_k}}$ by Eq. (5)

update $\widehat{\boldsymbol{\alpha}}_{\boldsymbol{G}_k}$ by Eq. (25)

end for

update $\boldsymbol{D}_{\boldsymbol{G}}^{t+1}$ by concatenating all $\boldsymbol{D}_{\boldsymbol{G}_k}$

update $\widehat{\boldsymbol{\alpha}}_{\boldsymbol{G}}^{t+1}$ by concatenating all $\widehat{\boldsymbol{\alpha}}_{\boldsymbol{G_k}}$

update $\boldsymbol{b}^{t+1}$ by Eq. (19)

end for

$\boldsymbol{u}^0 = \boldsymbol{D_G} * \widehat{\boldsymbol{\alpha}}_{\boldsymbol{G}}$

**Until** the stopping criterion is satisfied

## 3. Experimental Results

In this section, the results of extensive experiments are reported to validate the proposed method for few-view CT reconstruction. We also determine the impact of the number of best-matched patches $m$ and search window size $L$. In the experiments, three representative slices, abdominal, pelvic, and thoracic images, were tested to demonstrate the performance of the proposed GSR-SART. All the images were downloaded from the National Cancer Imaging Archive. In all the experiments, the image arrays are 20 ×20 cm and the system projection matrix, in fan-beam geometry, was obtained by Siddon's ray-driven algorithm [33] with 64 projection views evenly distributed over 360°. The distance from the source to the rotation centre is 40 cm and the distance from the detector centre to the rotation centre is 40 cm. We use a flat detector with 512 bins. The images are 256 ×256 pixels. All the experiments were performed in MATLAB 2017a on a PC equipped with an AMD Ryzen 5 1600 CPU at 3.2 GHz and 16 GB RAM.

### *3.1. Experimental results*

In this subsection, the experimental results of the different clinical images are given. The experiments were simulated under ideal conditions, which means that the measured projection data are noiseless. The parameters of GSR-SART were set as follows: the patch size was 8 × 8, which means that $P_s = 64$, $m$ was set to 40, and $L$ was set to 40. For the abdominal images, $\lambda = 1e-5$ and $\mu = 0.1$, for the pelvic image, $\lambda = 5e-5$ and $\mu = 0.1$, and for the thoracic image, $\lambda = 1.5e-5$ and $\mu = 0.08$. To further evaluate the proposed GSR-SART method, we compared GSR-SART with FBP, EM, SART, and TV-POCS. The peak signal-to-noise ratio (PSNR), root-mean-square error (RMSE) and structural similarity (SSIM) are utilized to quantitatively evaluate the performance of the methods.

The PSNR is defined as:

$$PSNR = 10 \times log_{10}\left(\frac{(\max(\boldsymbol{f}_i))^2}{(\sum_{i=1}^{N} (\boldsymbol{f}_i - \boldsymbol{f}_i^*)^2)/N}\right), \tag{27}$$

where $\boldsymbol{f}_i$ is the reconstructed value, $N$ is the size of $f$, and $\boldsymbol{f}_i^*$ is the golden reference value.

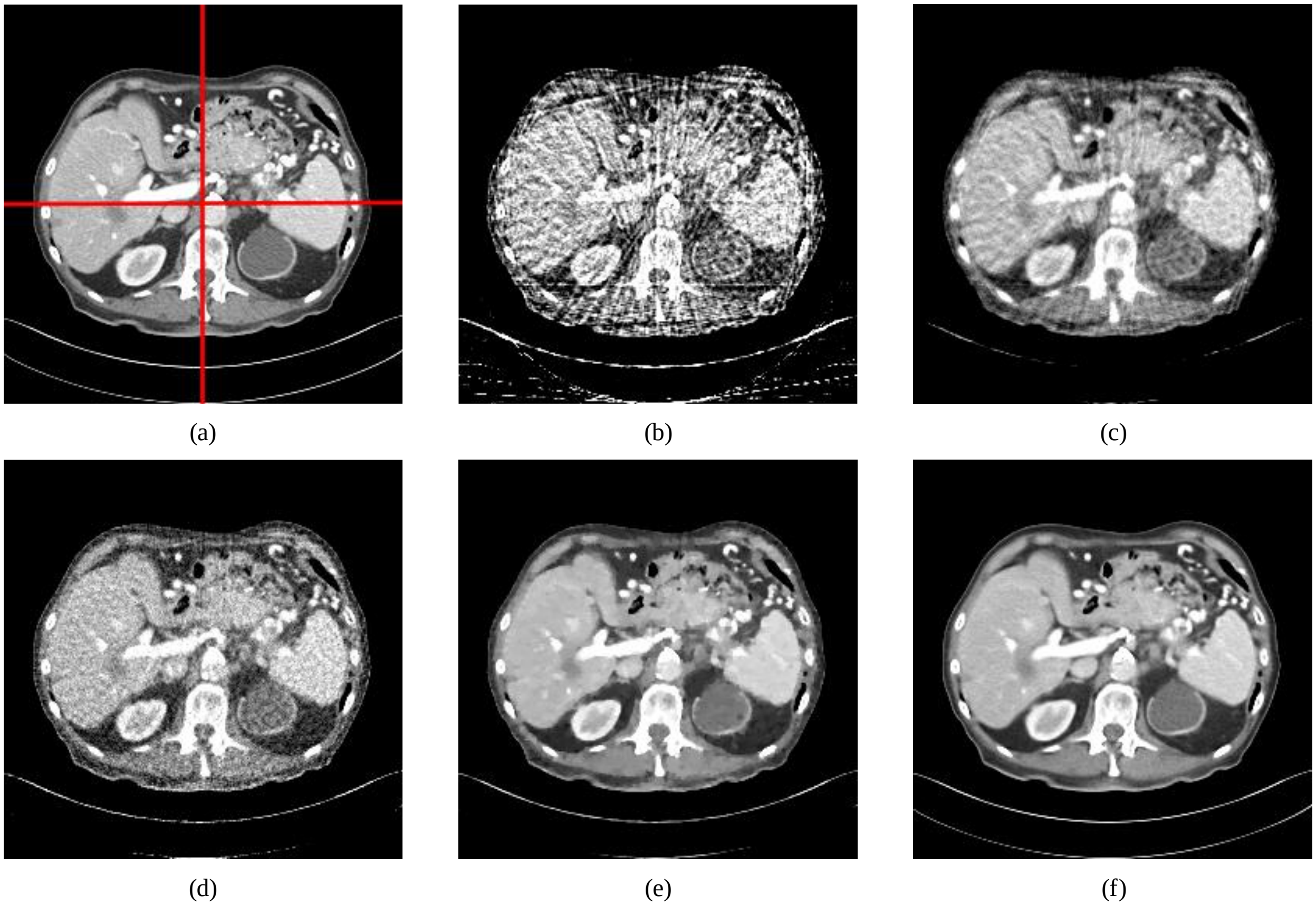

Fig. 2. Few-view reconstruction results of the abdominal image from 64 noiseless projections over 360°. The display window is [−150 250] HU. (a) Original image and results obtained by (b) FBP, (c) SART, (d) EM, (e) TV-POCS, and (f) GSR-SART.

The RMSE is defined as:

$$RMSE = \sqrt{\left(\sum_{i=1}^{N} (\boldsymbol{f}_i - \boldsymbol{f}_i^*)^2\right)/N}. \tag{28}$$

The SSIM is defined as:

$$\mathrm{SSIM}(\boldsymbol{f}, \boldsymbol{f}^*) = \frac{2\bar{\boldsymbol{f}}\bar{\boldsymbol{f}}^*(2\sigma_{\boldsymbol{f}\boldsymbol{f}^*}+c_2)}{(\bar{\boldsymbol{f}}^2+\bar{\boldsymbol{f}}^{*2}+c_1)(\sigma_{\boldsymbol{f}}^2+\sigma_{\boldsymbol{f}^*}^2+c_2)}, \tag{29}$$

where $\bar{\boldsymbol{f}}$ and $\bar{\boldsymbol{f}}^*$ are the mean values of $\boldsymbol{f}$ and $\boldsymbol{f}^*$, respectively, $\sigma_{\boldsymbol{f}\boldsymbol{f}^*}$ is the covariance of $\boldsymbol{f}$ and $\boldsymbol{f}^*$, and $c_1$ and $c_2$ are constants.

The original abdominal image and reconstruction results are shown in Fig. 2. In Fig. 2(b), the result of FBP contains severe streak artifacts due to the incomplete projection data. It can be observed in Figs. 2(c) and 2(d) that there are still undesirable artifacts in the SART and EM results. In Fig. 2(e), TV-POCS removes all the streak artifacts, but the reconstructed image of TV-POCS suffers from obvious over-smoothing effects. GSR-SART achieves the best visual effect in Fig. 2(f), which shows that it suppresses most of the artifacts without introducing any other side effects.

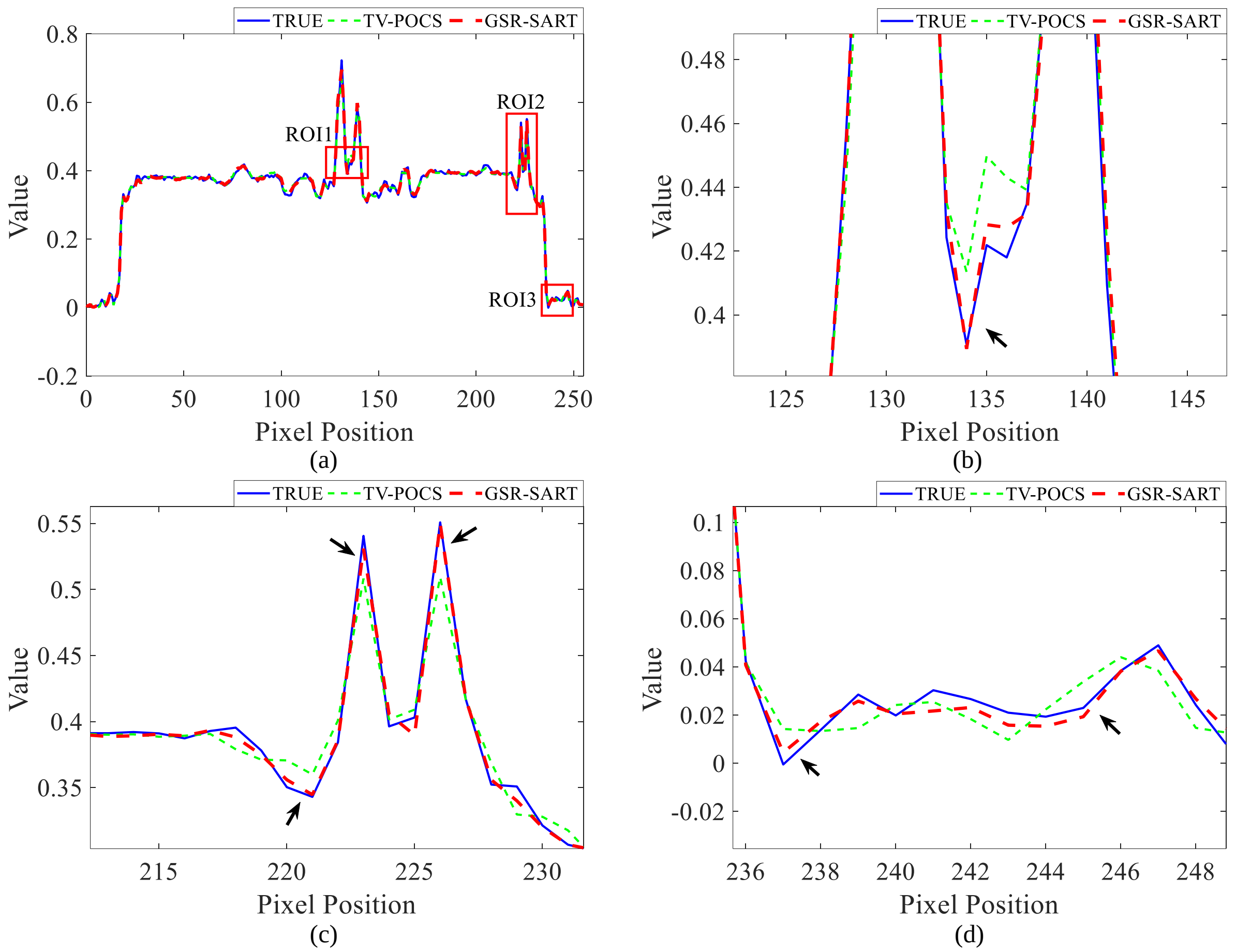

Fig. 3. Horizontal profiles (128th row) of the abdominal image reconstructed by different methods from 64 projection views. (a) Overall profiles and (b) ROI1, (c) ROI2, and (d) ROI3 of the overall profiles.

To further visualize the performance of the methods for the abdominal image, horizontal and vertical profiles of the abdominal image, which are indicated by red lines in Fig. 2(a), are shown in Figures 3 and 4. The profiles from the original image are given as references. Three regions of interest (ROIs) in each profile are enlarged for better visibility. Several arrows indicate the regions in which discrepancies can easily be identified. Here, the profiles generated from our algorithm are closer to the references. The quantitative evaluation of abdominal image is given in Table 1. It is obvious that our method achieves the best performance for all metrics, which demonstrates the ability of GSR-SART to better reduce artifacts and preserve structure than all the other methods.

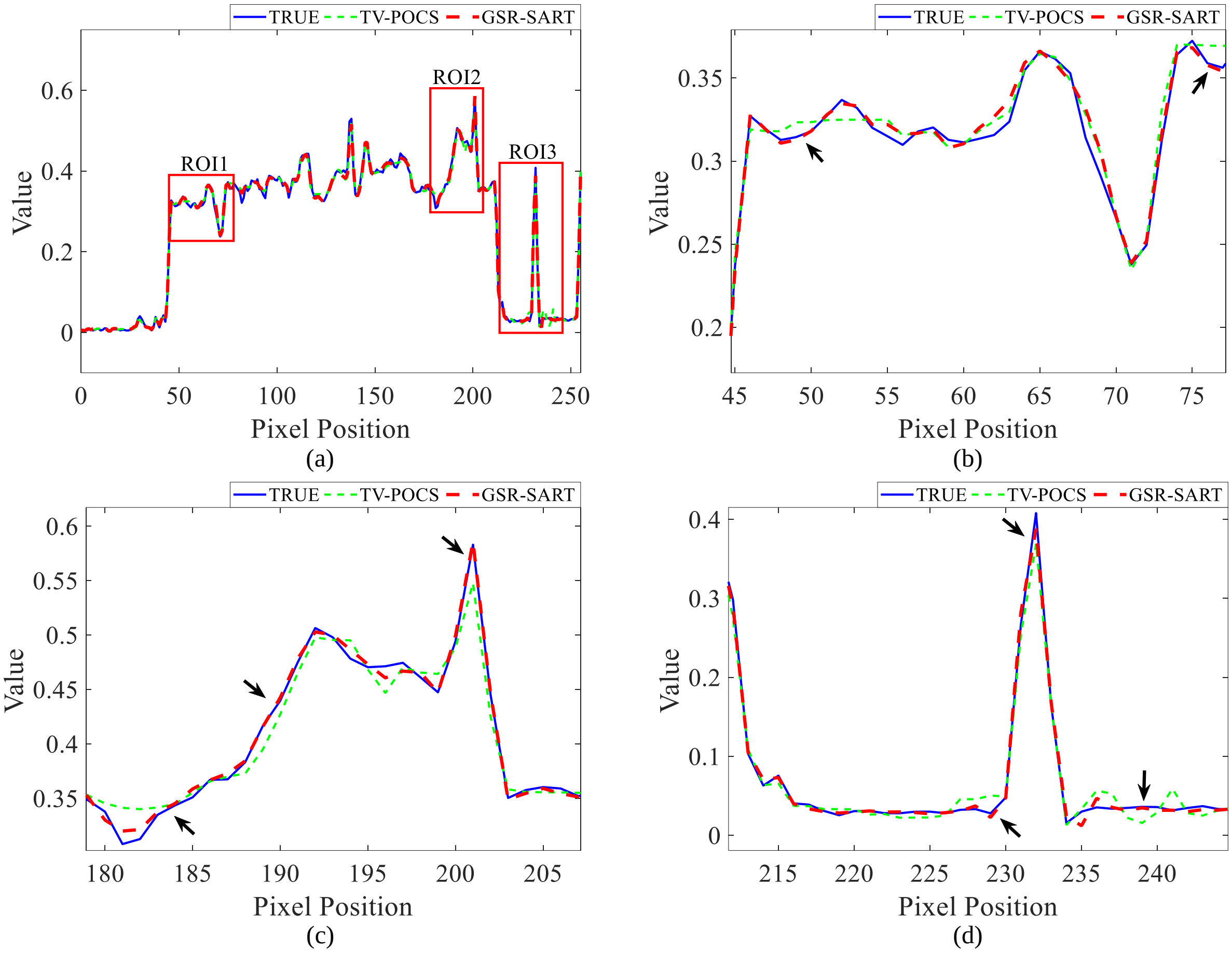

Fig. 4. Vertical profiles (128th column) of the abdominal image reconstructed by different methods from 64 projection views. (a) Overall profiles and (b) ROI1, (c) ROI2, and (d) ROI3 of the overall profiles.

In Fig. 5, the reconstruction results of the pelvic image are given. Because of the extremely low sampling ratio, it is difficult to obtain useful information from the result of FBP in Fig. 5(b). The SART and EM methods can only remove some of the streak artifacts. In Fig. 5(e), TV-POCS lowers the spatial resolution while eliminating the streak artifacts and the edges of the tissues are blurred to different degrees. Note that while GSR-SART eliminates most of the artifacts, it maintains the edges better than other methods in the region indicated by the red arrow.

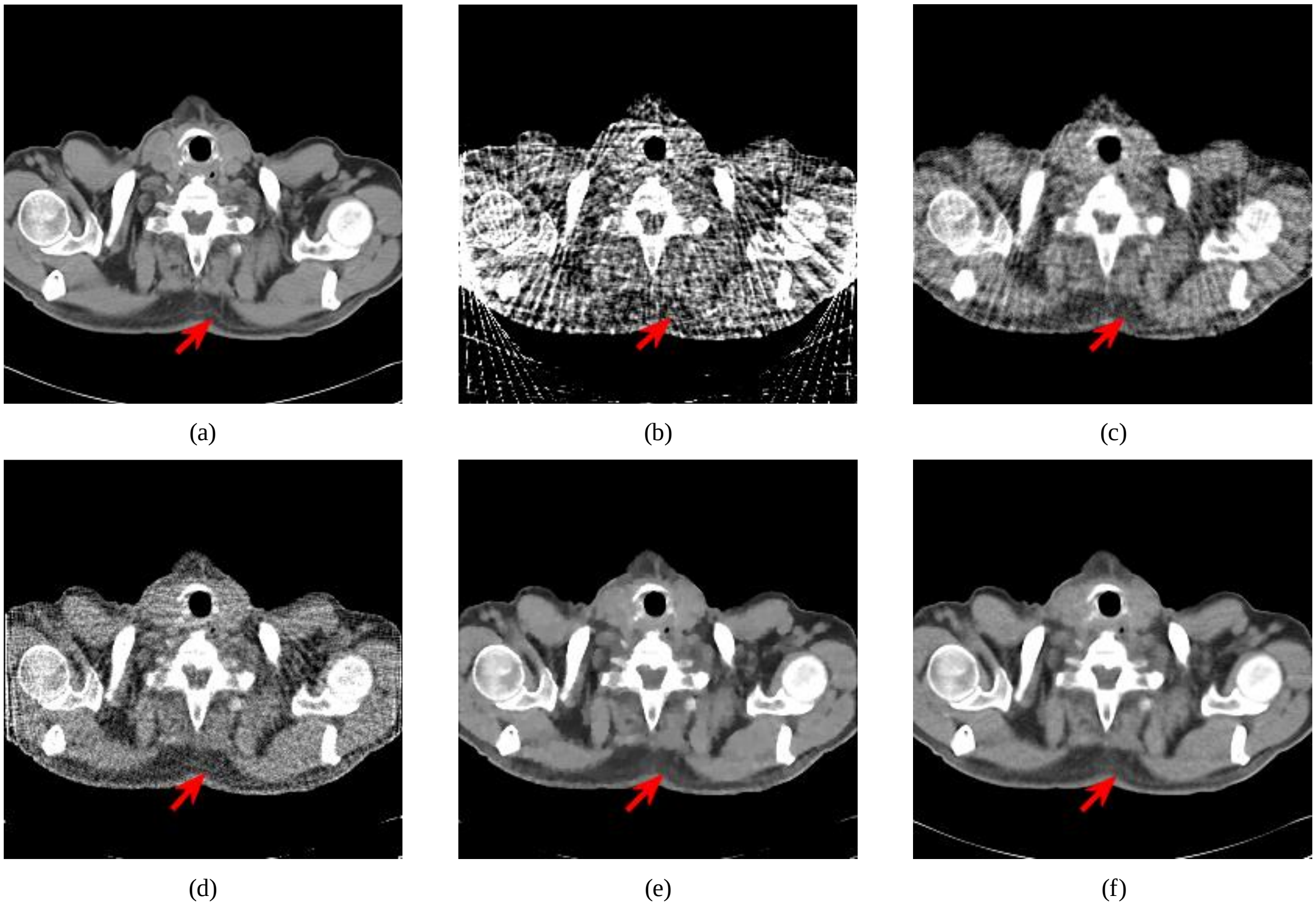

Fig. 5. Few-view reconstruction results of the pelvic image from 64 noiseless projections over 360°. The display window is [−150 250] HU. (a) Original image and results obtained by (b) FBP, (c) SART, (d) EM, (e) TV-POCS, and (f) GSR-SART.

To further demonstrate the performance of GSR-SART, the absolute difference images relative to the original images are shown in Fig. 6. Here, the loss of structural information in Fig. 6(b) is more than for other methods and the results from SART and EM still have artifacts. In Figs. 6(d) and 6(e), the artifacts are well suppressed and GSR-SART preserves more details, which can be observed in the lower part of the body, as indicated by the red arrow.

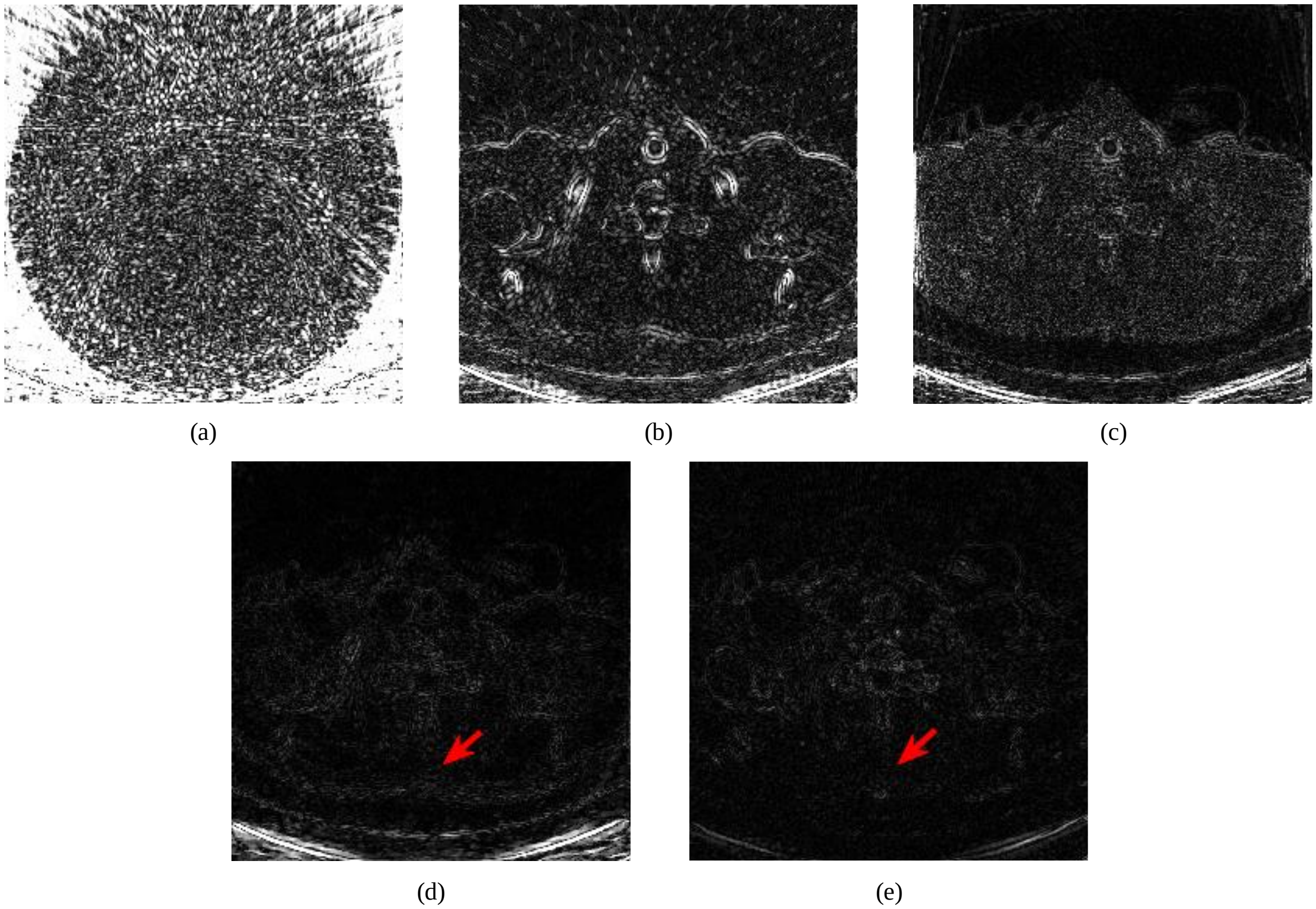

Fig. 6. Difference images relative to the original image. Results for (a) FBP, (b) SART (c) EM, (d) TV-POCS, and (e) GSR-SART.

The results of the thoracic image are shown in Fig. 7. In Fig. 7(b), the whole image is filled with streak artifacts and no clinically valuable structures can be recognized. Although SART and EM remove some artifacts, the spatial resolutions are not satisfactory and the blood vessels in the lungs are clearly blurred. TV-POCS and GSR-SART recover the most vessels in Figs. 7(e) and 7(f) and the spatial resolutions are close to the original image. However, artifacts still exist near the bones, as indicated by red arrows. The red square region of Fig. 7(a) is enlarged in Figure 8. It is easy to see that the artifacts and noise are severe and the spine is distorted heavily. In Fig. 8(e), the noise is still obvious and the structural details are blurred. Compared with other methods, GSR-SART suppresses more artifacts and noise and the edges of tissue are better maintained. The quantitative results are shown in Table 3. Consistent with the visual effects, GSR-SART has the best scores for all measurements and the improvements are impressive.

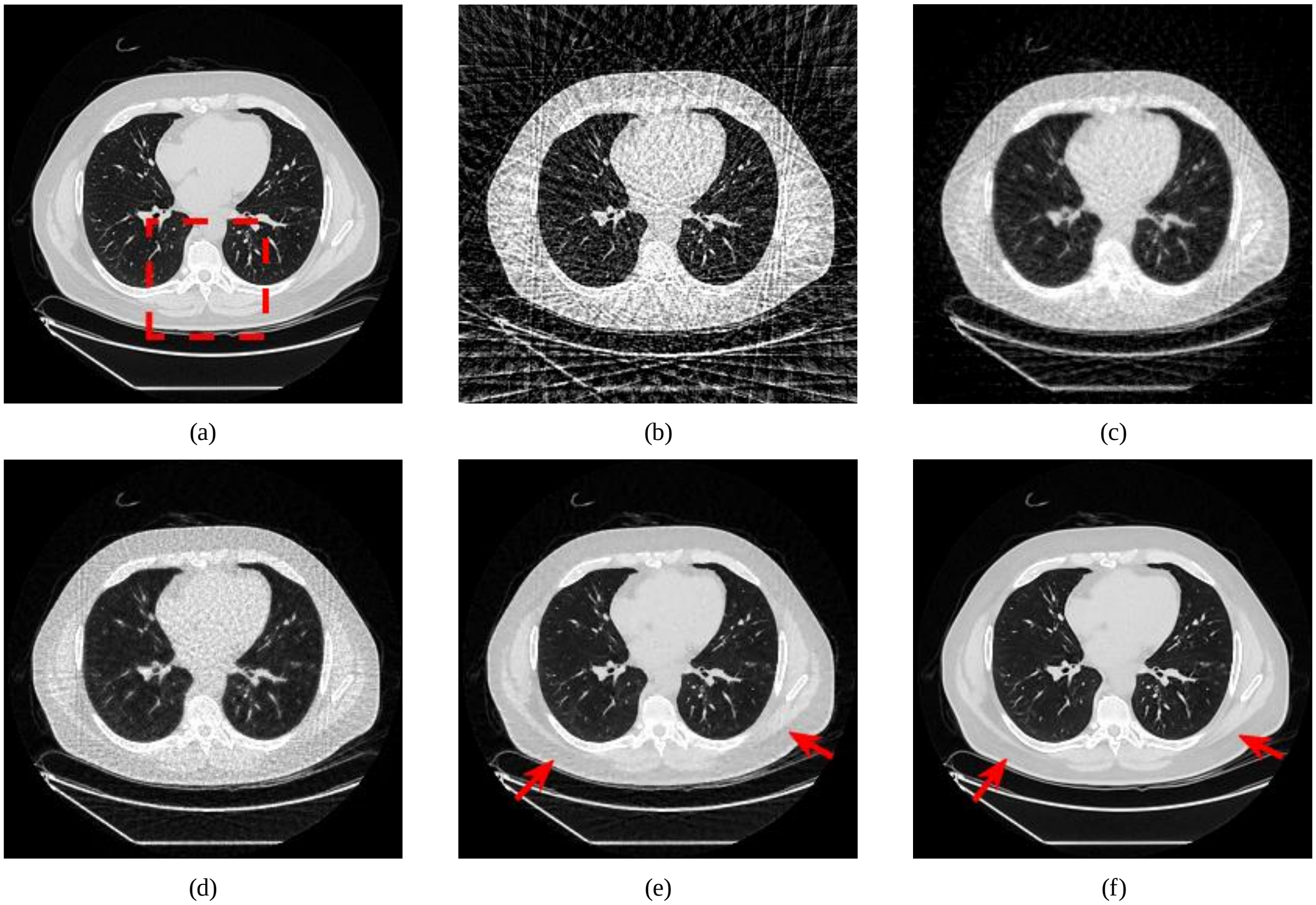

Fig. 7. Few-view reconstruction results of thoracic image from 64 noiseless projections over 360°. The display window is [-1000 250]HU. (a) Original image and results obtained by (b) FBP, (c) SART, (d) EM, (e) TV-POCS, and (f) GSR-SART.

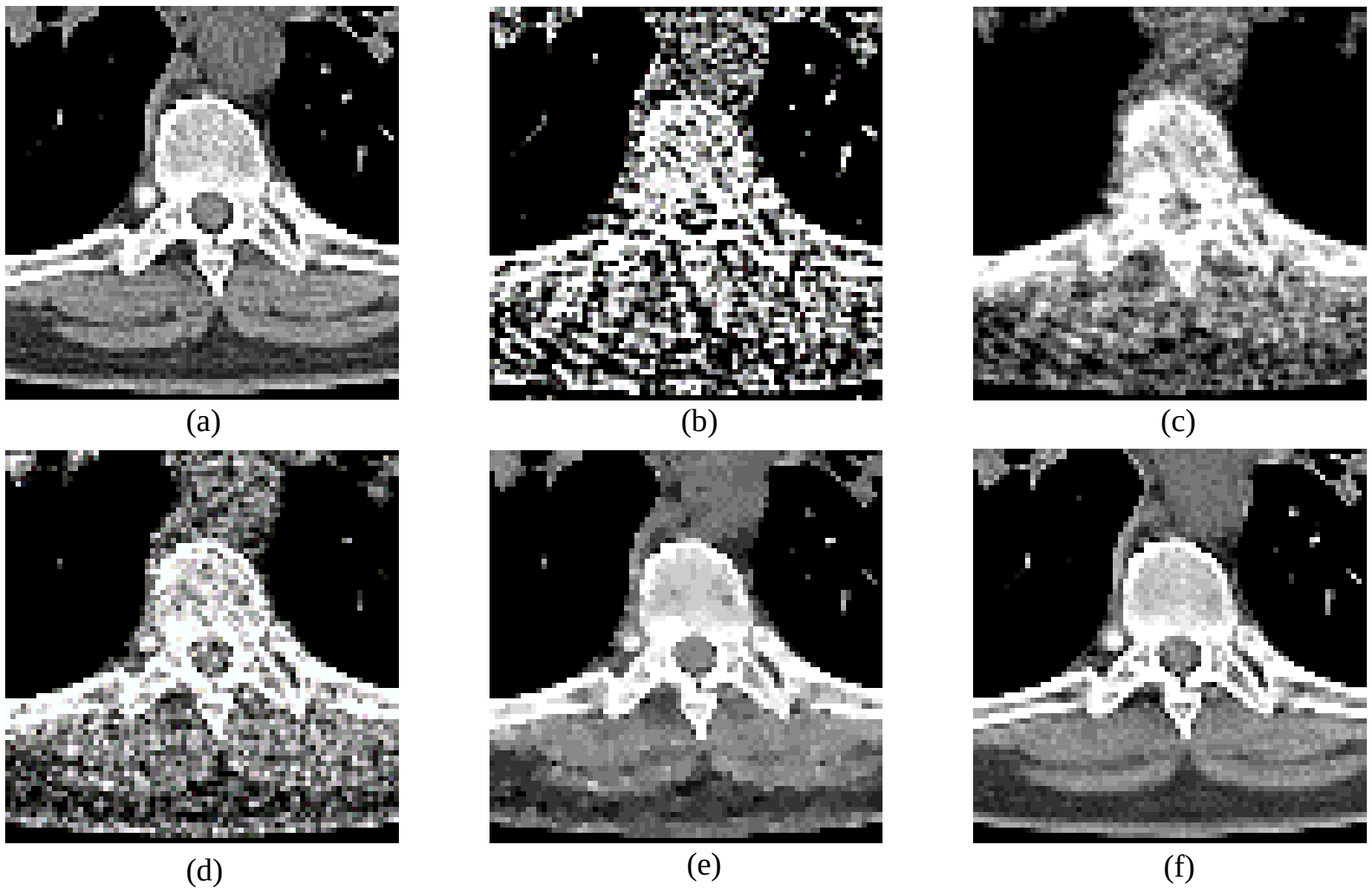

Fig. 8. Enlarged region of the thoracic image. The display window is [−150 250] HU. Enlarged regions of (a) the true image, (b) FBP, (c) SART, (d) EM, (e) TV-POCS, and (f) GSR-SART.

Table 1. Quantitative results obtained by different algorithms for the abdominal image

| | PSNR | RMSE | SSIM |
|---|---|---|---|
| FBP | 23.49 | 0.06690 | 0.51857 |
| SART | 31.05 | 0.02803 | 0.84241 |
| EM | 34.43 | 0.01900 | 0.90744 |
| TV-POCS | 36.95 | 0.01421 | 0.95726 |
| GSR-SART | **45.64** | **0.00522** | **0.98448** |

Table 2. Quantitative results obtained by different algorithms for the pelvic image

| | PSNR | RMSE | SSIM |
|---|---|---|---|
| FBP | 21.56 | 0.08358 | 0.47260 |
| SART | 32.02 | 0.02507 | 0.85824 |
| EM | 34.32 | 0.01924 | 0.91128 |
| TV-POCS | 37.32 | 0.01362 | 0.97062 |
| GSR-SART | **45.74** | **0.00516** | **0.98568** |

Table 3. Quantitative results obtained by different algorithms for the thoracic image

| | PSNR | RMSE | SSIM |
|---|---|---|---|
| FBP | 21.99 | 0.07949 | 0.34400 |
| SART | 30.45 | 0.03004 | 0.79577 |
| EM | 33.91 | 0.02015 | 0.87844 |
| TV-POCS | 38.28 | 0.01219 | 0.95866 |
| GSR-SART | **43.53** | **0.00666** | **0.98155** |

### *B. Parameter Selection*

#### *1) Number of Best Matched Patches $m$*

To investigate the sensitivity of $m$, experiments were performed with various $m$ ranging from 30 to 80 in steps of 10 with a fixed search window size of $40 \times 40$. The pelvic image was chosen as the test image. The results are shown in Fig. 9. It can be observed that the PSNR reaches peak at $m = 40$ and it slowly declines as $m$ increases further. In contrast to PSNR, the values of SSIM decrease monotonously as $m$ increases. One possible reason for this phenomenon is that PSNR focuses more on the reduction of artifacts and noise, which does not always correspond with structure preservation.

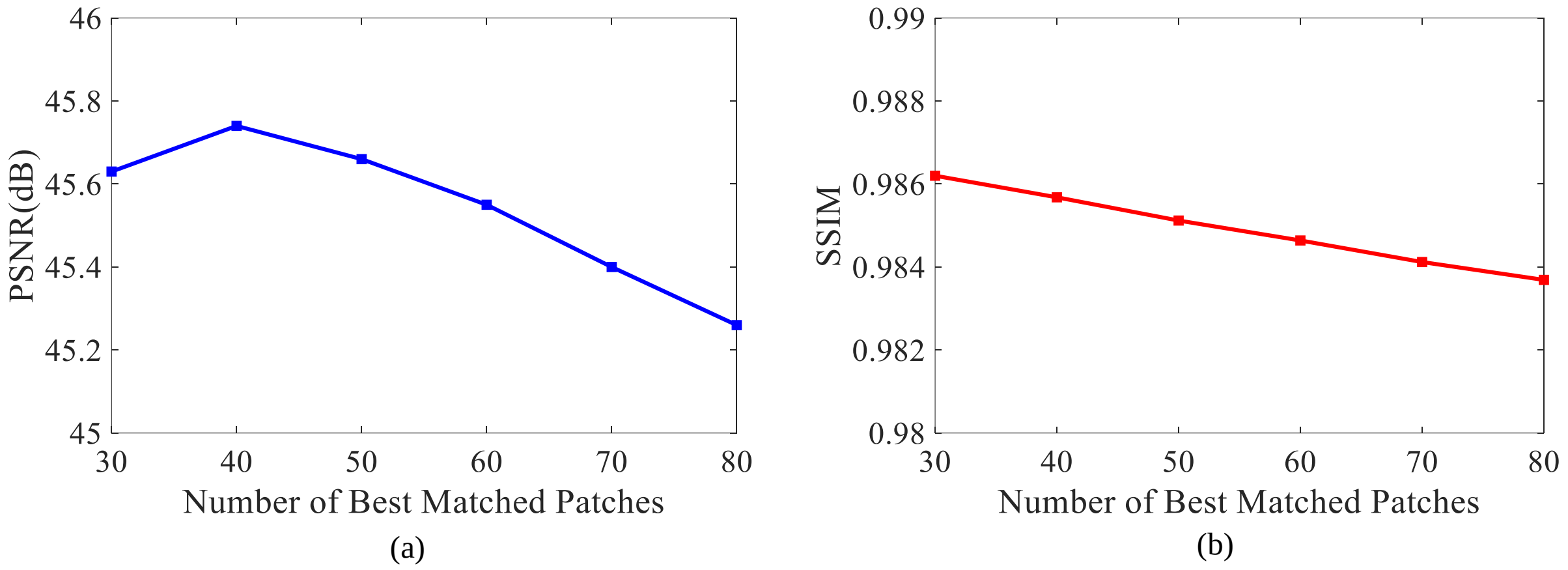

Fig. 9. Performance with respect to number of best-matched patches.

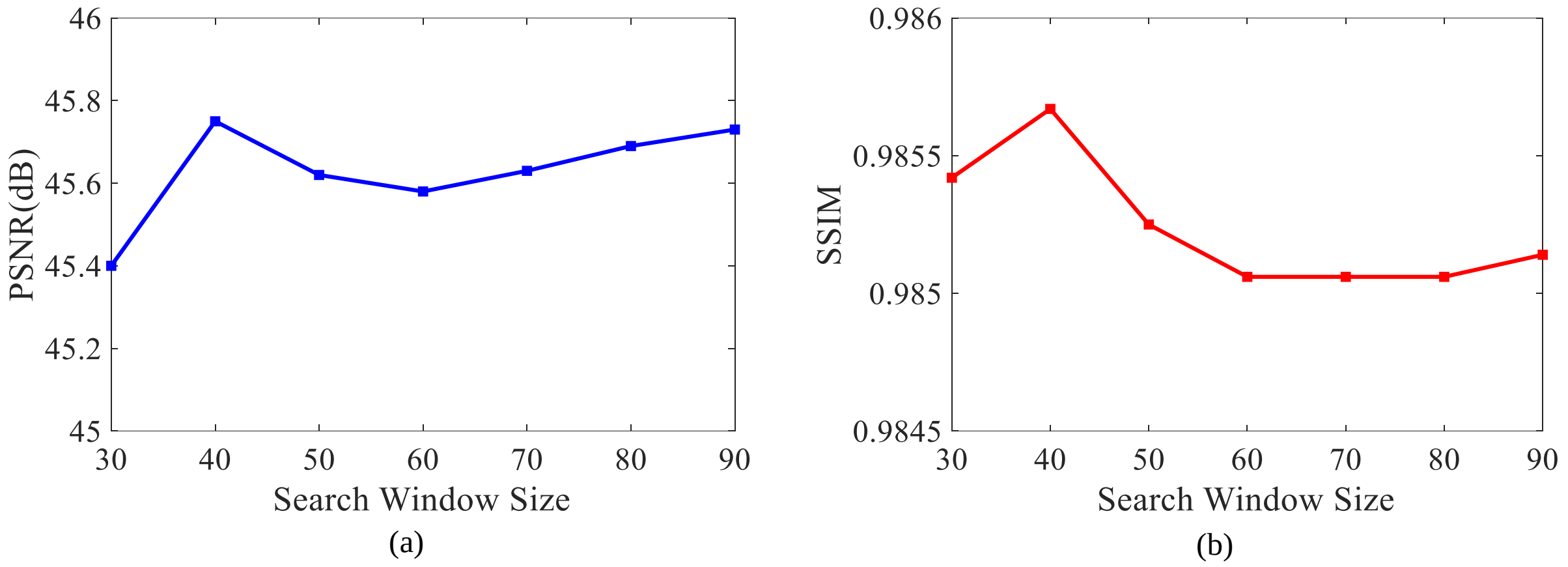

Fig. 10. Performance with respect to search window size.

*2) Search Window Size L*

To investigate the sensitivity of $L$, experiments were performed for various $L$ ranging from 30 to 90 in steps of 10 with a fixed number of best-matched patches of 40. The pelvic image was selected as the test image. The results are given in Fig. 10. Here, the values of PSNR and SSIM behave similarly. The values first increase and reach a peak when $L = 40$. After that, they decrease slowly. Although there is a rebound after 60, $L = 40$ is still the optimal selection.

## 4. Discussion and Conclusion

Despite the rapid development of CT imaging techniques, incomplete projection data reconstruction is still a major problem in this field. In this paper, we proposed a novel GSR-based SART algorithm for few-view CT reconstruction called GSR-SART. In this algorithm, we utilize a GSR model as the regularization term to eliminate streak artifacts and preserve structural details. To further explore nonlocal similarity in a target image, a dictionary of the GSR model, which is adaptively generated at each iteration, is learned from groups composed of similar patches. Three representative clinical slices of different parts of the human body were used to validate the performance of the proposed method. In all the results, our method performs better than all the other popular methods qualitatively and quantitatively under the same sampling conditions. Specifically, GSR-SART demonstrates a superior ability to reduce artifacts and preserve details.

In our experiments, the parameters were manually selected. Grid search is a reasonable method for parameter selection in simulations, but it is not practical in actual situations. We also observed that the optimal selection for each image was different, which makes this problem more complicated. A popular machine learning technique is a possible way to adaptively determine the optimal parameter set by learning from an external dataset.

Another problem we note here is the computational cost. Due to the introduction of group sparsity, the computational complexity of GSR-SART is heavier than that of original dictionary learning based methods. One possible solution to accelerate the computation is to implement a version that uses parallel computing. Distributed computing, computing clusters, and graphics processing units (GPUs) are three alternative approaches.

In conclusion, we are very encouraged by the promising performance of GSR with respect to artifact reduction and detail preservation for few-view CT. These results demonstrate the potential of the group-based sparse representation method for medical imaging. In the future, the proposed network framework will be refined and adapted to deal with other topics in CT imaging, such as low-dose CT, metal artifact reduction, and interior CT.

## **Acknowledgments**:

We thank Kim Moravec, PhD, from Edanz Group China (www.liwenbianji.cn/ac), for editing the English text of a draft of this manuscript. This work was supported in part by the National Natural Science Foundation of China under Grants 61671312 and 61302028 and National Key R&D Program of China under Grants 2017YFB0802300.